# Thermal spin injection and accumulation in CoFe/MgO tunnel contacts to *n*-type Si through Seebeck spin tunneling

Kun-Rok Jeon[1,2], Byoung-Chul Min[3], Seung-Young Park[4], Kyeong-Dong Lee[1,5], Hyon-Seok Song[1], Youn-Ho Park[3], and Sung-Chul Shin[1,6,*]

[1]*Department of Physics and Center for Nanospinics of Spintronic Materials, Korea Advanced Institute of Science and Technology (KAIST), Daejeon 305-701, Korea*

[2]*National Institute of Advanced Industrial Science and Technology (AIST), Spintronics Research Center, Tsukuba, Ibaraki 305-8568, Japan*

[3]*Center for Spintronics Research, Korea Institute of Science and Technology (KIST), Seoul 136-791, Korea*

[4]*Division of Materials Science, Korea Basic Science Institute (KBSI), Daejeon 305-764, Korea*

[5]*Department of Material Sciences and Engineering, Korea Advanced Institute of Science and Technology (KAIST), Daejeon 305-701, Korea*

[6]*Department of Emerging Materials Science, Daegu Gyeongbuk Institute of Science and Technology (DGIST), Daegu 711-873, Korea*

[*]e-mail: scshin@kaist.ac.kr; scshin@dgist.ac.kr

**We report the thermal spin injection and accumulation in crystalline CoFe/MgO tunnel contacts to *n*-type Si through Seebeck spin tunneling (SST). With the Joule heating (laser heating) of Si (CoFe), the thermally induced spin accumulation is detected by means of the Hanle effect for both polarities of the** **temperature gradient across the tunnel contact. The magnitude of the thermal spin signal scales linearly with the heating power and its sign is reversed as we invert the temperature gradient, demonstrating the major features of SST and thermal spin accumulation. We observe that, for the Si (CoFe) heating, the thermal spin signal induced by SST corresponds to the majority (minority) spin accumulation in the Si. Based on a quantitative comparison of thermal and electrical spin signals, it is noted that the thermal spin injection through SST can be a viable approach for the efficient injection of spin accumulation.**

Understanding the interplay between heat and spin transport is a fundamental and intriguing subject that also offers unique possibilities for emerging electronics based on the combination of thermoelectrics and spintronics[1-4]. Especially in semiconductor (SC) spintronics[5-8], where the injection, control, and detection of non-equilibrium spin populations (*i.e.* spin accumulation) in non-magnetic SCs are the main building blocks[5-9], the functional use of heat provides a new route to inject and control the spin accumulation in SCs without a charge current[4,7,8].

Recently, Le Breton *et al.*[10] introduced a conceptually new mechanism for the injection of spin accumulation ($\Delta\mu$), in which a temperature difference (or heat flow) across a ferromagnet (FM)/oxide/SC tunnel contact can induce $\Delta\mu$ into SC through Seebeck spin tunneling (SST). It was found that the SST effect, involving the thermal transfer of the spin angular momentum from FM to SC without a tunneling charge current, is a purely interface-related phenomenon of the ferromagnetic tunnel contact and the resultant $\Delta\mu$ is governed by the energy derivative of its tunnel spin polarization (TSP)[10]. As shown in a recent report[11], the SST and thermal spin accumulation ($\Delta\mu_{th}$) stem from the spin-dependent Seebeck coefficient of the ferromagnetic tunnel contact. It was also shown that the spin-dependent Seebeck coefficient gives rise to the tunnel magnetothermopower (or tunnel magneto Seebeck effect), *i.e.,* the dependence of the thermopower of a magnetic tunnel junction (MTJ) on the relative magnetic configuration of the two FM electrodes, as observed in MgO-based[12,13] and $Al_2O_3$-based[14] MTJs. Therefore, the SST effect offers a notable thermal approach for the generation of $\Delta\mu$ in SCs as well as for the functional use of heat in spintronic devices.

Whereas the SST and $\Delta\mu_{th}$ phenomena in *p*-type Si have been intensively studied using a $Ni_{80}Fe_{20}/Al_2O_3$ tunnel contact[10], the demonstration of these phenomena in *n*-type Si with different tunnel contact materials has not been realized thus far. Furthermore, considering the fact that the sign and magnitude of the induced $\Delta\mu_{th}$ are crucially dependent on the energy derivative of TSP of the ferromagnetic tunnel interface[10,11], the investigation of those in the different material system is highly desirable for a complete understanding of the SST mechanism and the successful implementation of its functionality in SC spintronics.

Here, we report the achievement of thermal spin injection and accumulation in crystalline CoFe/MgO contacts to *n*-type Si through SST. Using the Joule heating (laser heating) of Si (CoFe) and Hanle measurements, we explicitly demonstrate the major features of SST and thermal spin accumulation that the magnitude of thermally injected spin signal scales linearly with the heating power and its sign is reversed when the temperature gradient across the tunnel contact is reversed. It is also observed that, for the Si (CoFe) heating, the thermal spin signal induced by the SST corresponds to the majority (minority) spin accumulation in the Si. Based on a quantitative comparison of the thermal and electrical spin signals, the thermal spin injection through SST is suggested as an effective route to inject the spin accumulation.

Figure 1 schematically illustrates the device geometry and measurement scheme used in the present study. We fabricated a device consisting of highly (001) textured $Co_{70}Fe_{30}$(5 nm)/MgO(2 nm) tunnel contacts to *n*-Si(001) in which the *n*-Si channel is heavily As-doped ($n_d \sim 2.5 \times 10^{19}$ cm$^{-3}$ at 300 K)[15]. A Si-active channel with a lateral dimension of 300×500 μm$^2$ and three identical tunnel contacts (*a-c*, 100×300 μm$^2$) were defined by standard photolithographic and Ar-ion beam etching techniques. These contacts are separated by about 100 μm from each other, which is much longer than the spin-diffusion length. The magnetic easy axis of the $Co_{70}Fe_{30}$ (hereafter abbreviated as CoFe) contacts are along the [110] direction of Si parallel to the long axes of the contacts. For electrical isolation (at the sides of the tunnel contacts) and for the contact pads, approximately 120-nm-thick $Ta_2O_5$(115 nm)/$Al_2O_3$(2 nm) layers and 110-nm-thick Au(100 nm)/Ti(10 nm) layers, respectively, were grown by a sputtering technique. Details of the sample preparation as well as the structural and electrical characterization are available in the literature[15].

To detect thermally-induced spin accumulation through the SST process, we generated the temperature difference ($\Delta T \equiv T_{Si} - T_{CoFe}$) across the tunnel contact *a* employing two different heating methods (Joule heating and laser heating), which are the identical to those used in a previous study[16]. For the SC heating (Fig. 1(a)), we applied a heating current ($I_{heating}$) through the Si channel using two contacts *b* and *c*, which causes Joule heating and raises $T_{Si}$ with respect to $T_{CoFe}$ ($\Delta T > 0$). For the FM heating (Fig. 1(b)), the Au bond pad was heated using a laser beam with a wavelength of 532 nm and a maximum power of 200 mW. The laser beam spot (with a diameter of 5-10 μm and a skin depth of ~3 nm) on the 100-nm-thick Au pad is located approximately 300 μm from one edge of the tunnel contact (note that the size of the Au pad is large enough and the position of the laser beam spot is far enough away to prevent the direct illumination of the Si layer). Hence, a part of the heat generated from the laser beam passes through the contact, resulting in $T_{CoFe}>T_{Si}$ ($\Delta T<0$).

In an open-circuit geometry, where the tunneling charge current ($I_{tunnel}$) is zero, the measured voltage between the contact *a* and *d* is given by $V=V_{th}+\Delta V_{TH}$[10,11]. The first term $V_{th}$ is the thermovoltage maintaining the zero net charge current ($I_{tunnel}=0$) and the second term $\Delta V_{TH}$ is the SST voltage due to the induced $\Delta\mu_{th}$ in the SC. The $\Delta V_{TH}$ can be detected by means of the Hanle effect[17,18]. When we apply a magnetic field ($B_z$) transverse to the spins in the SC, the induced $\Delta\mu_{th}$ is suppressed via spin precession and dephasing. This results in a voltage change ($\Delta V_{TH}$), directly proportional to $\Delta\mu_{th}$, with a Lorentzian line shape as a function of $B_z$. As noted in recent reports[19,20], the induced $\Delta\mu_{th}$ can also be detected with an in-plane magnetic field ($B_x$) parallel to the film plane, giving rise to the inverted Hanle effect. Both the normal and inverted Hanle effects arising respectively from the suppression and recovery of the spin accumulation are indicative of the presence of spin accumulation. The full spin accumulation is given by the sum of the normal and inverted Hanle signals. Therefore, the two measurement schemes (Fig. 1) using the normal and inverted Hanle effects[17-20] provide a concrete means of demonstrating the SST and resultant $\Delta\mu_{th}$ in the SC.

Figure 2 shows the measured voltage changes ($\Delta V_{TH} \propto \Delta\mu_{th}$) under perpendicular ($B_z$) and in-plane ($B_x$) magnetic fields corresponding to the normal and inverted Hanle effects, respectively, while heating the Si side ($\Delta T>0$) as described above. As shown in the $\Delta V_{TH}$–$B_z$ plots of Figs. 2(a) and 2(b), large normal Hanle signals ($\Delta V_{TH, normal}$) with a Lorentzian line shape were observed at 300 K (base *T*). The inverted Hanle signals ($\Delta V_{TH, inverted}$)[19,20] in $B_x$ (Figs. 2(c) and 2(d)), roughly 1.3-1.4 times larger than the magnitude of $\Delta V_{TH, normal}$, were also clearly measured. It is worth noting that the line shapes of Hanle curves ($\Delta V_{TH}$) and the ratio of $\Delta V_{TH, inverted}$ to $\Delta V_{TH, normal}$ are consistent with those of electrical Hanle signals ($\Delta V_{EH}$) obtained by electrical spin injection/extraction using the same contact $a$[21]. The observed thermal Hanle signals ($\Delta V_{TH, normal}$, $\Delta V_{TH, inverted}$) show identical curve irrespective of the polarity in $I_{heating}$, indicating that the sign and magnitude of the induced $\Delta \mu_{th}$ in the Si are identical for both heating currents.

Figures 2(e)-2(h) summarize the magnitude of the thermal Hanle signals ($\Delta V_{TH, normal}$, $\Delta V_{TH, inverted}$) as a function of $I_{heating}$ (up to ±10 mA). These figures clearly show that $\Delta V_{TH, normal}$ and $\Delta V_{TH, inverted}$ scale quadratically (linearly) with $I_{heating}$ ($I_{heating}^2$). These results strongly support that the observed Hanle signals mainly come from the thermally driven spin accumulation in Si, which scales linearly with the $\Delta T$[10,11].

We performed a decisive test[10] using the Hanle measurements in larger perpendicular magnetic fields ($B_z$) (see Fig. 1) to rule out any possibility that the observed spin accumulation is induced by the origin of the spin caloric effect (e.g., the spin Nernst effect)[4] in Si. It is possible that the spin Nernst effect[4] produces a spin current (along the $z$-axis) transverse to a thermal gradient (in the $x$- or $y$-axis direction) via the spin orbit interaction of Si (see Fig. 1). This would cause spin accumulation at the Si interface with the spins pointing parallel to the film plane (note that the direction of the accumulated spins is transverse to the thermal gradient and to the spin current). Figure 3(a) illustrates the expected characteristic behavior[10] of the $\Delta \mu_{th}$ signal as a function of $B_z$ for two different origins (($i$) and ($ii$)) of $\Delta \mu_{th}$. ($i$) If the spins were accumulated by origin of the spin Nernst effect in Si without thermal spin injection from FM to Si, they would be unrelated to the magnetization ($M$) of FM[10]. This gives rise to a simple reduction of $\Delta\mu_{th}$ signal in small values of $B_z$ (the blue line in Fig. 3(a)) because the direction of the induced spins is always orthogonal to $B_z$ irrespective of the $M$ of FM. (*ii*) In contrast, if the spins induced in Si were thermally injected from FM, their direction would coincide with that of the $M$ of FM[10,19,20] even with larger values of $B_z$. This results in distinctly different behavior of the $\Delta\mu_{th}$ signal with larger $B_z$ values (the red line in Fig. 3(a)) as compared to the case (*i*). As $B_z$ is increased, the $\Delta\mu_{th}$ value is sharply reduced due to the Hanle effect; thereafter, it gradually increases when the $M$ of FM rotates out of the film plane. When the $M$ and the induced spins in the Si are fully aligned with $B_z$ higher than $\mu_0 M_s$ of FM (where $\mu_0$ is the permeability of the free space and $M_s$ is the saturation magnetization), the $\Delta\mu_{th}$ signal eventually becomes saturated.

Figure 3(b) shows the obtained thermal Hanle signals ($\Delta V_{TH, normal}$, $\Delta V_{TH, inverted}$; normalized) under applied magnetic fields ($B_z$, $B_x$) up to 50 kOe with an $I_{heating}$ value of ±10 mA (note that each Hanle curve is the average of two curves measured with positive and negative $I_{heating}$ values). The $\Delta V_{TH, normal}$ signal (closed symbol) clearly reveals the distinctive behavior with $B_z$; the Hanle effect in a small $B_z$ (<2 kOe), the increase of $\Delta V_{TH, normal}$ due to the rotation of $M$ in an intermediate $B_z$ (2-22 kOe), and the saturation of $\Delta V_{TH, normal}$ in a large $B_z$ (>22 kOe). This result excludes any possible origin for thermal spin accumulation in our system that does not involve spin injection from CoFe to Si. It was noted earlier that the difference of the $\Delta V_{TH, normal}$ value in the saturation region from the peak value at zero $B_z$ is associated with the initial spin precession and dephasing effect caused by the local magnetostatic fields due to the finite roughness of the CoFe/MgO interface, as proven by the inverted Hanle signal ($\Delta V_{TH, inverted}$, open symbol) under an in-plane magnetic field ($B_x$)[19,20].

The sign of $\Delta\mu_{th}$ can be determined by a direct comparison[10,19] with that of electrical Hanle signals ($\Delta V_{EH, normal}$, $\Delta V_{EH, inverted}$) obtained from three-terminal Hanle (TTH) measurements[22-24] for the same contact $a$[21]. The sign of $\Delta V_{EH, normal}$ in $B_z$ is negative for a negative electrical voltage ($V_{el}$) ($V_{CoFe}-V_{Si}<0$, electron spin injection) and positive for a positive $V_{el}$ ($V_{CoFe}-V_{Si}>0$, electron spin extraction). As shown in Figs. 2(e) and 2(g), the sign of $\Delta V_{TH, normal}$ (in $B_z$) is identical to that of the former $\Delta V_{EH, normal}$ ($V_{el}<0$), indicating that $\Delta\mu_{th}$ produced by the thermal spin injection with Si heating ($T_{Si}>T_{CoFe}$) has the same sign as $\Delta\mu_{el}$ induced by electrical spin injection. Given the positive TSP of bcc (Co)Fe/MgO(001) tunnel interfaces[25,26], the induced $\Delta\mu_{th}$ by SST when $T_{Si}>T_{CoFe}$ corresponds to the majority spin accumulation ($\Delta\mu>0$) in the Si, where a larger number of electrons have spin angular momentum parallel to the direction of the $M$ of FM. Based on SST theory[10,11], the positive sign of $\Delta\mu_{th}$ (or majority spin accumulation) for the SC heating indicates that the sign of the energy-derivative of the TSP of the CoFe/MgO tunnel interface at the Fermi-level ($E_F$) is negative. This corresponds to the case in which the TSP value of the CoFe/MgO interface varies weakly for energies below $E_F$ but decays significantly for energies above $E_F$. It should also be noted that the same sign of the experimental results ($\Delta\mu>0$ for SC heating) was observed in $n$-type Ge with $Co_{70}Fe_{30}$/MgO[19] and $Co_{60}Fe_{20}B_{20}$/MgO[27] contacts and in $p$-type Si with a $Ni_{80}Fe_{20}$/$Al_2O_3$ contact[10].

We estimated that the associated $\Delta\mu_{th}$ in the CoFe/MgO tunnel contact to $n$-type Si is (+)0.37 meV with the maximum $I_{heating}$ (±10 mA) through the typical conversion formula of $\Delta\mu_{th}=(-2e)\Delta V_{TH}/\gamma$[10,11,22-24] based on the existing theory[9,28]. In this calculation, we used the total measured $\Delta V_{TH}$ of (−)0.13 mV, given by the sum of $\Delta V_{TH, normal}$ and $\Delta V_{TH, inverted}$, and the assumed TSP ($\gamma$) of (+)0.7 for a crystalline CoFe/MgO tunnel interface[24,25].

Another major feature of the SST and $\Delta\mu_{th}$ is that, for a given energy-dependence of TSP of the ferromagnetic tunnel contact, the sign of the thermal spin signal is reversed when $\Delta T$ is reversed[10,11]. In order to demonstrate this, we employed a laser beam to heat the FM instead of Joule heating[16]. This approach provides a simple way to heat the FM ($T_{CoFe}>T_{Si}$, $\Delta T<0$) effectively and minimize the contribution of spurious effects such as current-in-plane (CIP) tunneling and anisotropic magnetoresistance (AMR) on the Hanle signal.

Figures 4(a) and 4(b) show the measured voltage signals ($V$) as a function of perpendicular ($B_z$, Fig. 4(a)) and in-plane ($B_x$, Fig. 4(b)) magnetic fields while varying the incident laser power ($P_{inc.\ laser}$) from 0 to 100 mW at 300 K (base $T$). At zero $P_{inc.\ laser}$ (black symbols), no magnetic field dependence of the signal is detected, as expected. With an increase of $P_{inc.\ laser}$ from 20 to 100 mW (purple to red symbols), clear normal and inverted Hanle signals are observed and the amplitudes of both Hanle signals gradually increase. As depicted in the $\Delta V_{TH}$–$P_{inc.\ laser}$ plots (Figs. 4(c) and 4(d)), the obtained Hanle signals ($\Delta V_{TH,\ normal}$, $\Delta V_{TH,\ inverted}$) scale almost linearly with $P_{inc.\ laser}$, indicating that the obtained signals stem from the thermal spin injection and accumulation in Si.

More importantly, the observed sign of $\Delta V_{TH}$ is obviously reversed with respect to the Si heating case. The $\Delta V_{TH,\ normal}$ ($\Delta V_{TH,\ inverted}$) value is positive (negative) for $T_{CoFe}>T_{Si}$ (see Figs. 4(c) and 4(d)) whereas it is negative (positive) for $T_{CoFe}<T_{Si}$ (see Figs. 2(e) and 2(f)). This result clearly demonstrates another key feature of SST and $\Delta\mu_{th}$ that the thermal spin signal is reversed when $\Delta T$ is reversed[10,11].

In addition, an analysis based on the sign of Hanle signals allows us to exclude another possible origin of the spin signal, such as the spin-polarized hot-electron injection via conventional Seebeck effect, in the laser-heating experiment. Thermal excitation by laser heating can produce an electron flow in the Au pad and the CoFe layer. Nevertheless, this hot-electron flow cannot be an origin of the obtained Hanle signals, since the observed sign of $\Delta V_{\text{TH, normal}}$ (a positive sign, minority spin accumulation; see Fig. 4(c)) with the laser heating of CoFe is opposite to the expected sign of $\Delta V_{\text{EH, normal}}$ (a negative sign, majority spin accumulation) due to the injection of the spin-polarized hot-electron into Si, whose transmission in FM is known to be larger for majority spins than for minority spins[29]. This strongly suggests that the obtained Hanle signals with laser heating do not arise from the hot-electron flow in the ferromagnetic tunnel contact.

For a quantitative analysis and for a comparison study, we plotted the magnitude of the Hanle signals ($\Delta V_{\text{TH}}$, $\Delta V_{\text{EH}}$) obtained in the same contact $a$ (Fig. 1) caused by the thermal and electrical spin accumulation (see Figs. 4(a) and S(c)) as a function of the driving term. The thermal driving term is $S_0 \Delta T (=V_{\text{th}})$[11], which should be compared to the electrical driving term of $R_{\text{contact}} I_{\text{tunnel}} (=V_{\text{el}})$[11]. Here, $S_0$ is the charge thermopower (or charge Seebeck coefficient), $R_{\text{contact}}$ is the contact resistance, and $I_{\text{tunnel}}$ is the tunneling current across the contact. Figures 5(a) and 5(b), respectively, show the $\Delta V_{\text{TH, normal}}$–$V_{\text{th, eff}}$ and $\Delta V_{\text{EH, normal}}$–$V_{\text{el}}$ plots. It is noted that, because the reference contact $d$ in our measurement scheme is located far away from the ferromagnetic tunnel contact $a$ (see Fig. 1), the offset voltage (or $V_{\text{B=0}}$–$\Delta V_{\text{TH}}$) in Fig. 4(a), which mainly arises from the Seebeck effect across the MgO tunnel barrier and partly from the Seebeck effects inside the Au and $n$-Si electrodes, should be considered as an effective value for the $V_{\text{th}}$ (or $V_{\text{th,}}$

$_{\text{eff}}$).

A noteworthy aspect of these plots is that the thermal spin accumulation (red symbol in Fig. 5(a)) requires about 100 times smaller value of $V_{\text{th, eff}}$ to obtain a similar magnitude of $\Delta V_{\text{Hanle, normal}}$ compared to the electrical spin accumulation case (blue symbol in Fig. 5(b)). This means that the proportionality factor between the spin accumulation and the driving term for the thermal spin injection is much larger than that for the electrical spin injection. According to the model[11], the proportionality factor of the electrical spin injection is limited by the absolute value of TSP, which cannot be larger than one (or 100 %) by definition. However, such a restriction does not exist for the proportionality factor of the thermal spin injection, which is governed by the energy derivative of TSP, since there is no limit for the energy derivative of TSP (note that, in principle, the proportionality factor for thermal spin accumulation can be arbitrarily large for suitably engineered ferromagnetic tunnel contacts)[11]. Therefore, the experimental result supports the theoretical proposition[11] that, for a ferromagnetic tunnel contact with large energy-derivative of TSP and proper thermal interface resistance, the thermal spin injection through SST can be an effective way to inject spin accumulation into SCs.

In conclusion, we have achieved the thermal spin injection and accumulation in crystalline CoFe/MgO contacts to *n*-type Si through SST. With the Joule heating (laser heating) of Si (CoFe) and subsequent Hanle measurements, we have demonstrated the major features of SST and thermal spin accumulation that the magnitude of a thermally injected spin signal scales linearly with the heating power and its sign is reversed as the temperature gradient across the tunnel contact is inverted. A quantitative analysis and a comparison study of the thermal and electrical spin signals suggest that the thermal spin injection through SST is a viable approach for the efficient injection of the spin accumulation.

This work was supported by the DGIST R&D Program of the Ministry of Education, Science and Technology of Korea (11-IT-01); by the KIST institutional program (2E22732 and 2V02720) and by the Pioneer Research Center Program (2011-0027905); and by the KBSI grant No.T33517 for S-YP. The authors would like to thank H. Saito and R. Jansen at AIST for the high-field measurements by PPMS and for the helpful discussions. One of the authors (K-RJ) acknowledges a JSPS Postdoctoral Fellowship for Foreign Researchers.

**Figure captions**

FIG. 1. (Color online) Schematic illustration of the device geometry and measurement scheme for Seebeck spin tunneling and thermal spin accumulation with (a) the Joule heating of Si and (b) the laser heating of CoFe.

FIG. 2. (Color online) (a)-(d) Obtained thermal Hanle signals ($\Delta V_{\text{TH, normal}}$, $\Delta V_{\text{TH, inverted}}$) under applied magnetic fields ($B_z$, closed circles; $B_x$, open circles) for heating currents ($I_{\text{heating}}$) of ±5 and ±10 mA at 300 K (base $T$), in the case of Si heating ($\Delta T > 0$). (a)/(b)

and (c)/(d) represent $\Delta V_{\text{TH, normal}}$ in $B_z$ and $\Delta V_{\text{TH, inverted}}$ in $B_x$ for an $I_{\text{heating}}$ value of $\pm 5/\pm 10$ mA, respectively. (e)-(h) Summary of the magnitudes of $\Delta V_{\text{TH, normal}}$ and $\Delta V_{\text{TH, inverted}}$ as a function of $I_{\text{heating}}$ ($I_{\text{heating}}^2$). (e) $\Delta V_{\text{TH, normal}}$ and (f) $\Delta V_{\text{TH, inverted}}$ with $I_{\text{heating}}$, together with a quadratic fit. (g) $\Delta V_{\text{TH, normal}}$ and (h) $\Delta V_{\text{TH, inverted}}$ with $I_{\text{heating}}^2$, together with a linear fit.

FIG. 3. (Color online) (a) Expected characteristic behavior of the thermal spin accumulation ($\Delta \mu_{\text{th}}$) signal as a function of the perpendicular magnetic field ($B_z$) for two different origins ((*i*) and (*ii*)) of $\Delta \mu_{\text{th}}$: (*i*) $\Delta \mu_{\text{th}}$ by thermal spin injection from CoFe to Si (red line) and (*ii*) $\Delta \mu_{\text{th}}$ without thermal spin injection from FM to Si (blue line). (b) Obtained thermal Hanle signals ($\Delta V_{\text{TH, normal}}$, $\Delta V_{\text{TH, inverted}}$; normalized) under applied magnetic fields ($B_z$, $B_x$) up to 50 kOe with the heating current ($I_{\text{heating}}$) of $\pm 10$ mA. Note that each Hanle curve is the average of two curves measured with positive and negative $I_{\text{heating}}$ values.

FIG. 4. (Color online) (a) Obtained thermal spin signals ($\Delta V_{\text{TH, normal}}$, $\Delta V_{\text{TH, inverted}}$) under (a) perpendicular ($B_z$) and (b) in-plane ($B_x$) magnetic fields as a function of the incident laser power ($P_{\text{inc. laser}}$) at 300 K (base $T$) in the case of CoFe heating ($\Delta T < 0$). (c) $\Delta V_{\text{TH, normal}}$ and (d) $\Delta V_{\text{TH, inverted}}$ as a function of $P_{\text{inc. laser}}$, together with a linear fit.

FIG. 5. (Color online) (a) Hanle signal ($\Delta V_{\text{TH}}$) obtained by thermal spin accumulation with the laser heating of CoFe (see Fig. 4(a)) as a function of the effective thermovoltage ($V_{\text{th, eff}}$), defined as $V_{B=0} - \Delta V_{\text{TH}}$. (b) Hanle signal ($\Delta V_{\text{TH}}$) obtained by electrical spin accumulation using the same contact *a* (see Fig. S(c)) as a function of the electrical voltage ($V_{el}$), defined as $V_{B=0} - \Delta V_{EH}$.

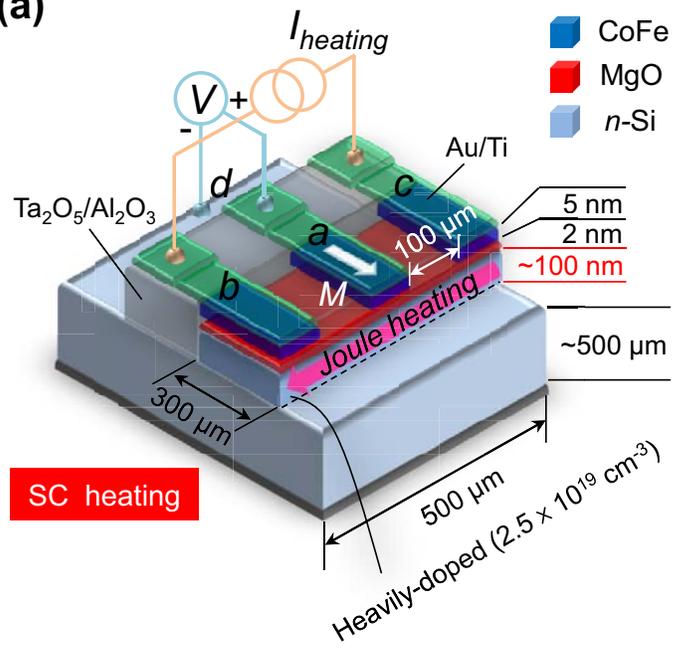 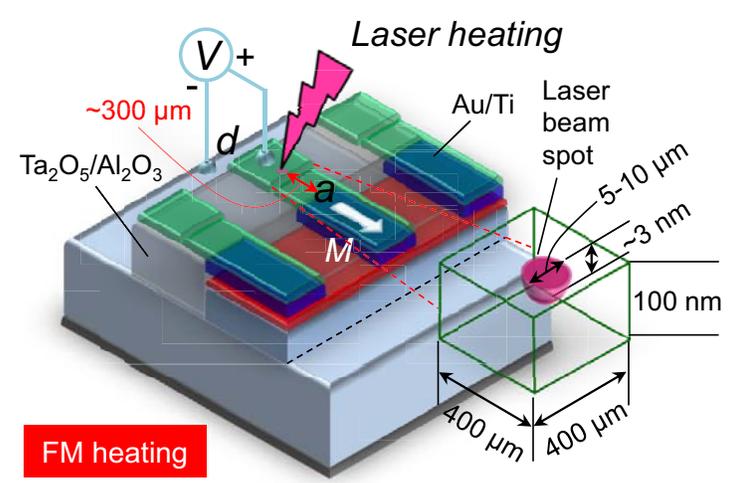

Fig. 1

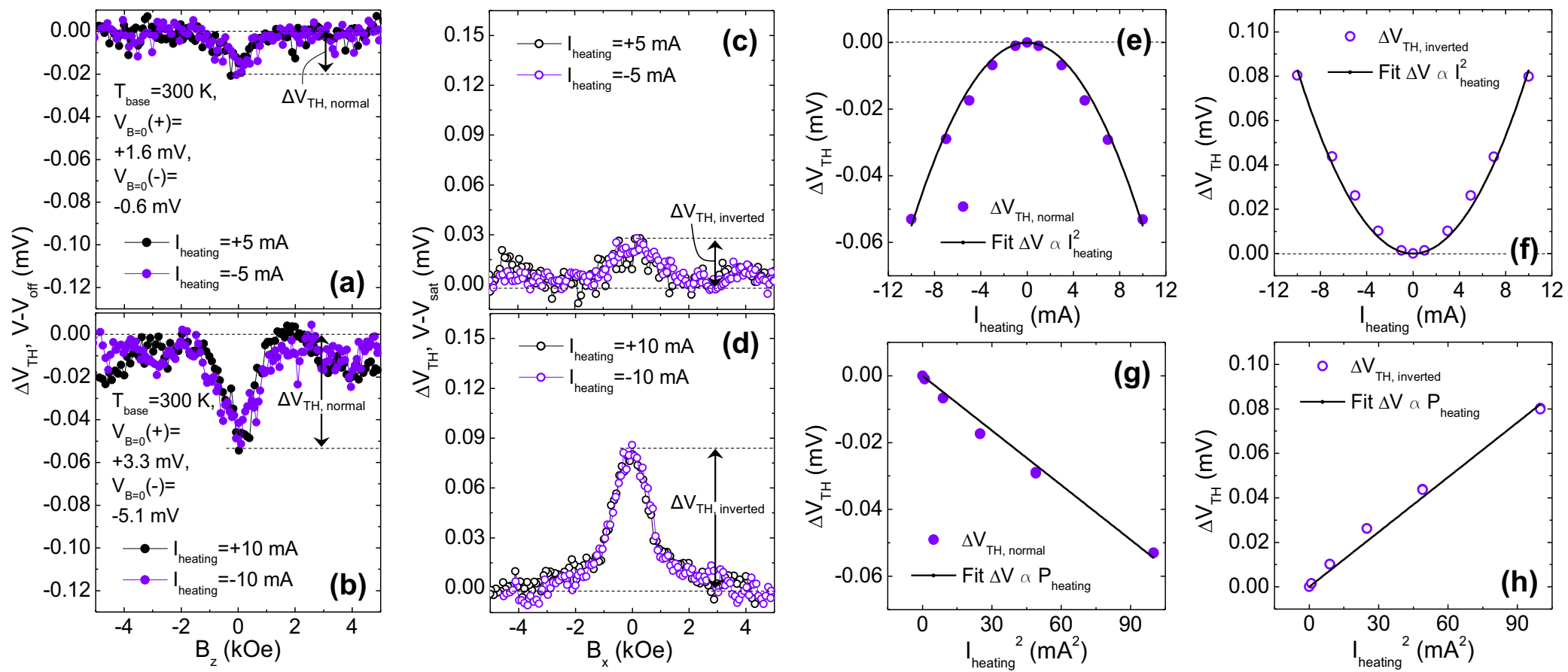

Fig. 2

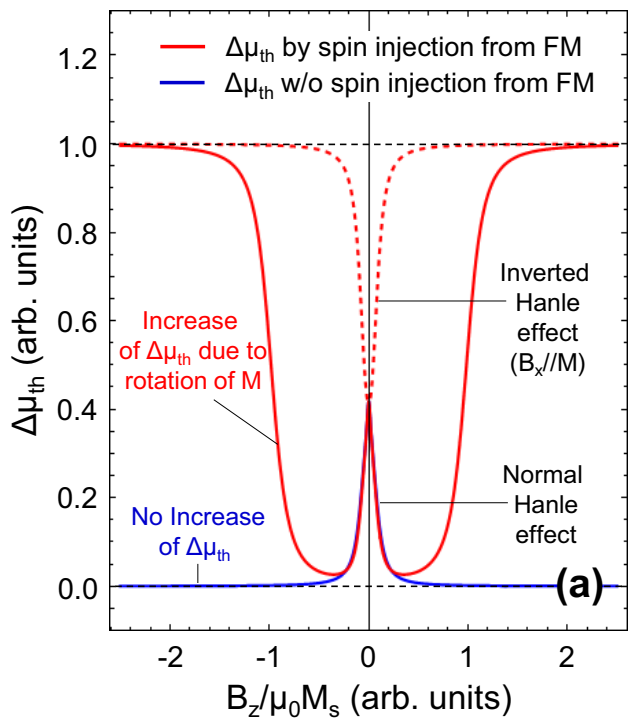 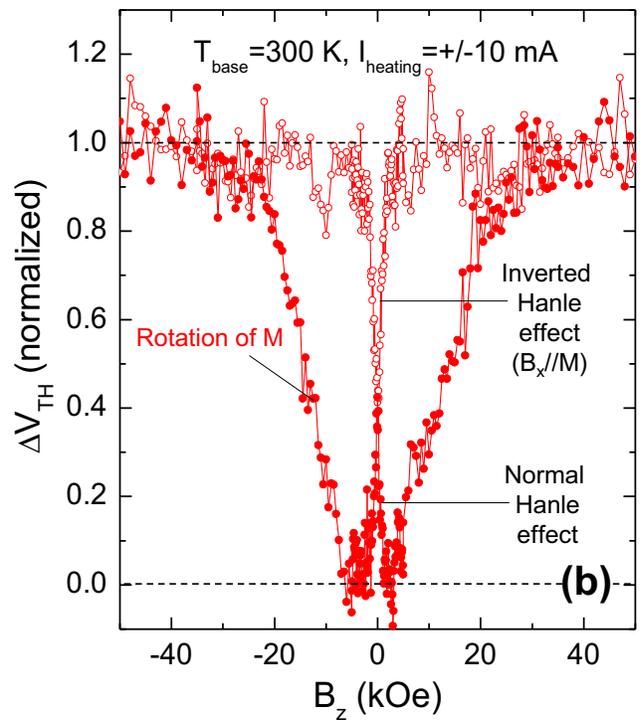

Fig. 3

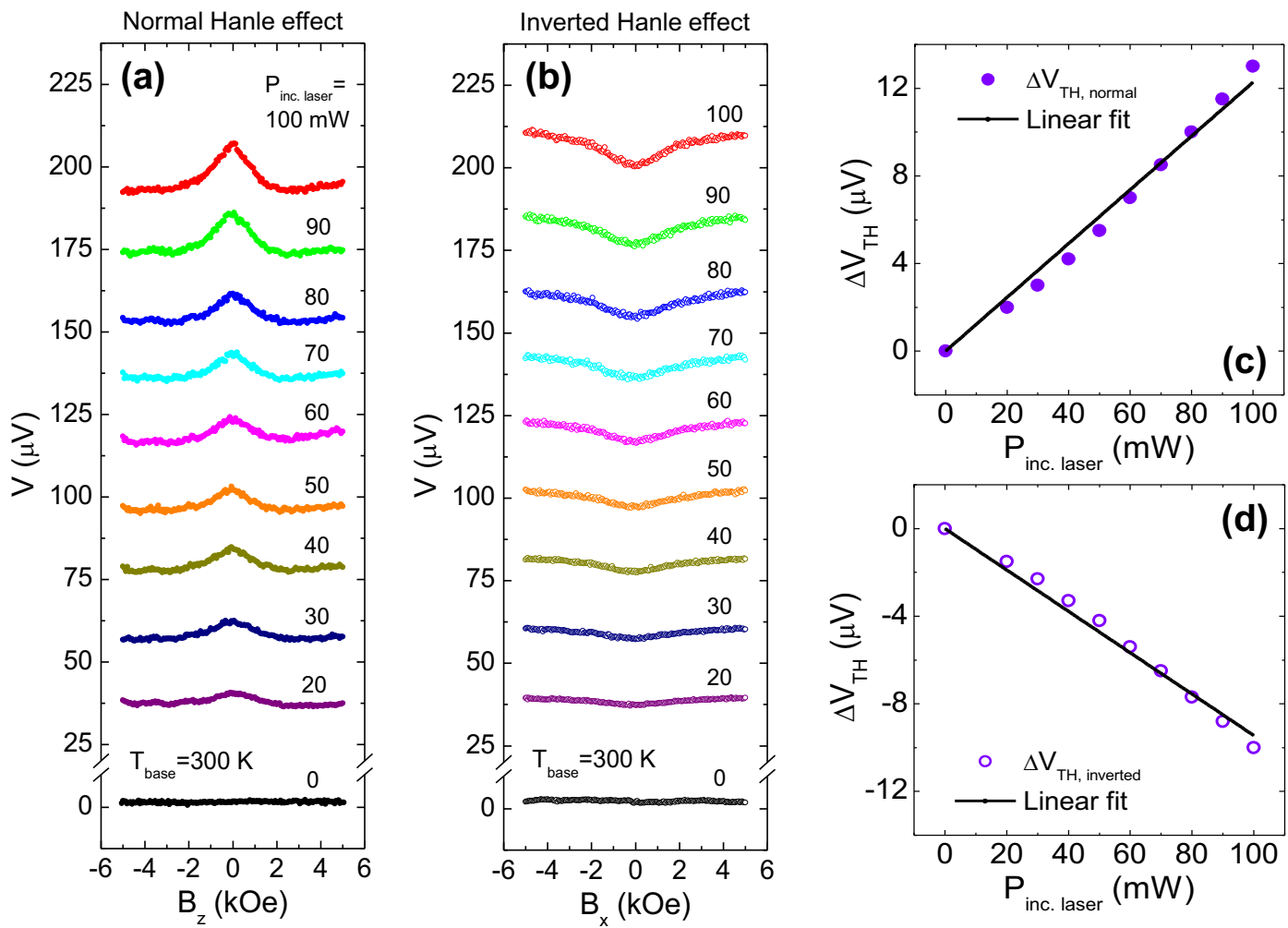

Fig. 4

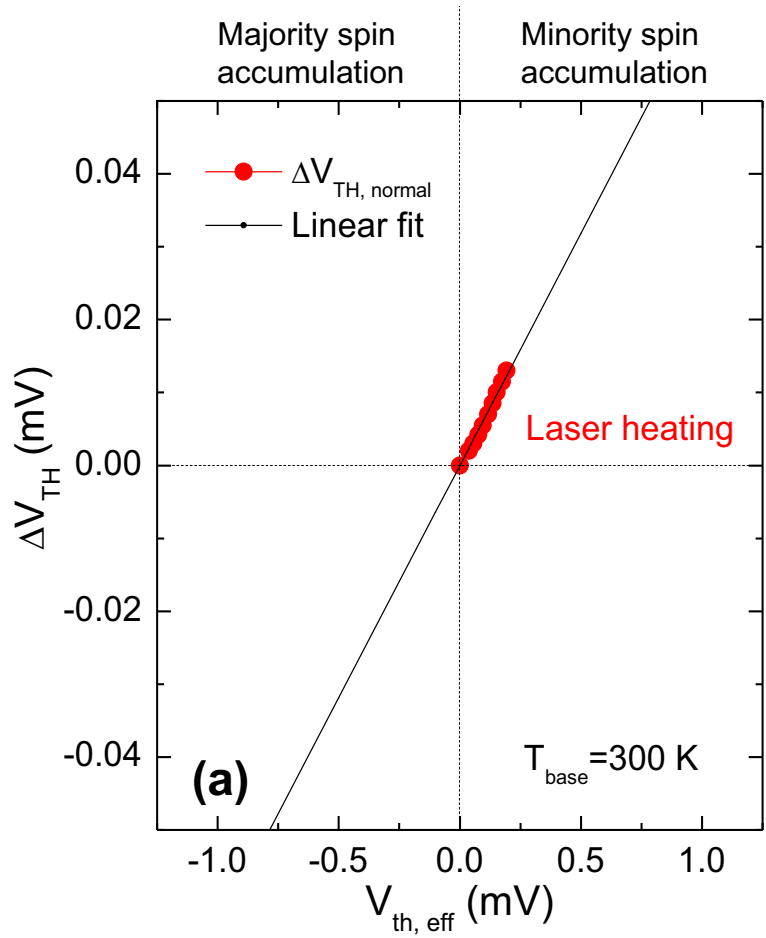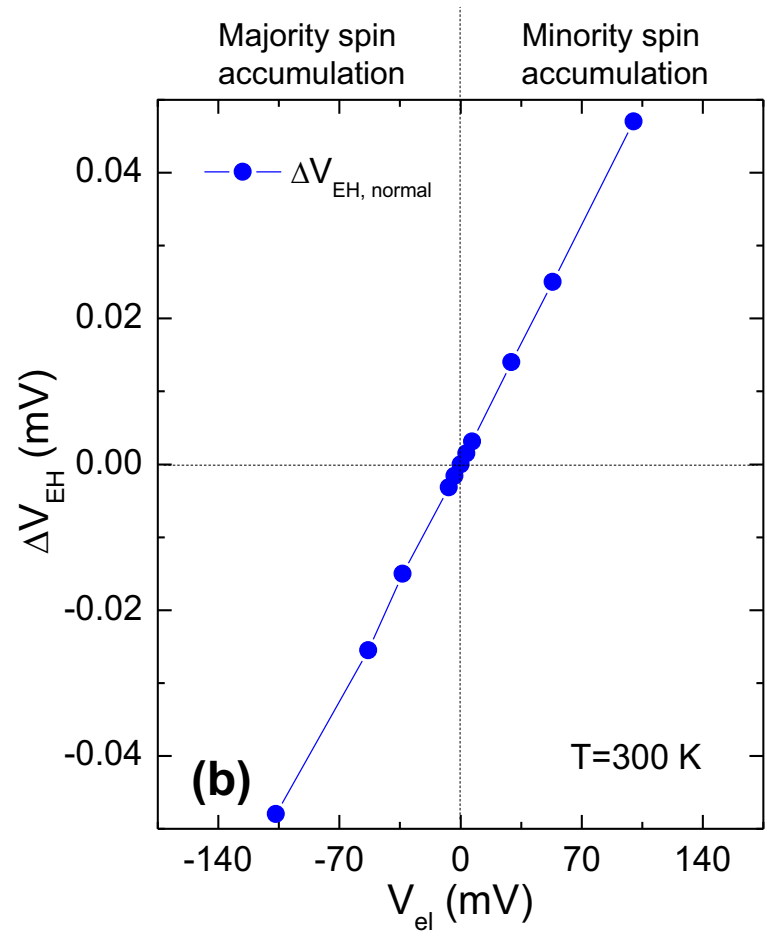

Fig. 5